\documentclass[aps,pra,preprint,twocolumn,superscriptaddress,longbibliography,10pt]{revtex4-1} 

\usepackage{amsmath,amssymb}
\usepackage{graphicx}
\usepackage{epsfig}
\usepackage{color}

\begin{document}

\title{Fast compressive Raman bio-imaging via matrix completion}

\author{Fernando Soldevila}
\address{GROC-UJI, Institute of New Imaging Technologies (INIT), Universitat Jaume I, E12071 Castell\`o, Spain}
\author{Jonathan Dong}
\address{Laboratoire Kastler Brossel, CNRS UMR 8552, Ecole Normale Sup\'{e}́rieure, PSL Research University Sorbonne Universit\'{e}́s \& Universit\'{e}́ ́ Pierre et Marie Curie Paris 06, F-75005, Paris, France}
\address{Laboratoire de Physique Statistique, CNRS, PSL Universit\'{e}s \& Ecole Normale Sup\'{e}rieure,
Sorbonne Universit\'{e}s et Universit\'{e} Pierre et Marie Curie, 75005, Paris, France}
\author{Enrique Tajahuerce}
\address{GROC-UJI, Institute of New Imaging Technologies (INIT), Universitat Jaume I, E12071 Castell\`o, Spain}
\author{Sylvain Gigan}
\address{Laboratoire Kastler Brossel, CNRS UMR 8552, Ecole Normale Sup\'{e}́rieure, PSL Research University Sorbonne Universit\'{e}́s \& Universit\'{e}́ ́ Pierre et Marie Curie Paris 06, F-75005, Paris, France}
\author{Hilton B. de Aguiar}
\email{h.aguiar@phys.ens.fr}
\address{D\'{e}́partement de Physique, Ecole Normale Sup\'{e}́rieure/PSL Research University, CNRS, 24 rue Lhomond, 75005 Paris, France}

\begin{abstract}
Raman microscopy is a powerful method combining non-invasiveness with no special sample preparation. Because of this remarkable simplicity, it has been widely exploited in many fields, ranging from life and materials sciences, to engineering. Notoriously, due to the required imaging speeds for bio-imaging, it has remained a challenge how to use this technique for dynamic and large-scale imaging. Recently, compressive Raman has been put forward, allowing for fast imaging, therefore solving the issue of speed. Yet, due to the need of strong a priori information of the species forming the hyperspectrum, it has remained elusive how to apply this technique for microspectroscopy of (dynamic) biological tissues. Combining an original spectral under-sampling measurement technique with matrix completion framework for reconstruction, we demonstrate fast and inexpensive label-free molecular imaging of biological specimens (brain tissues and single cells). Therefore, our results open interesting perspectives for clinical and cell biology applications using the much faster compressive Raman framework.
\end{abstract}

\maketitle 

\begin{figure*}[ht]
    \centering
    \includegraphics[width=2\columnwidth]{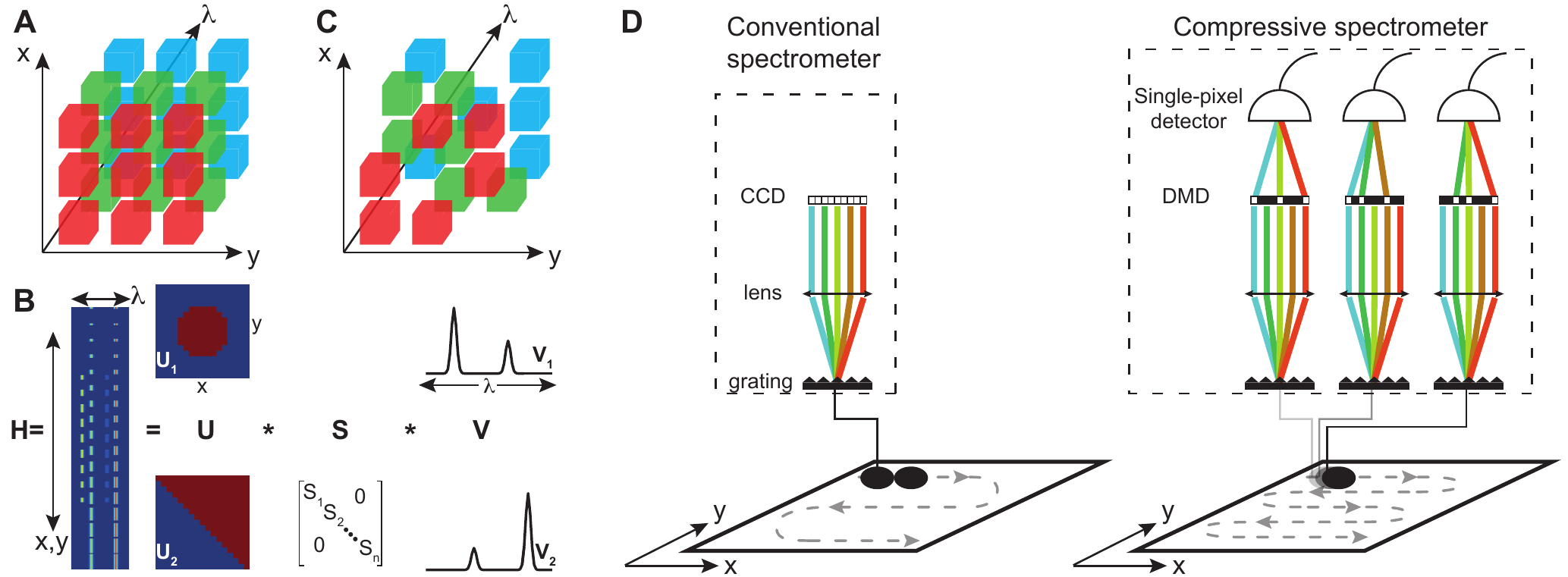}
    \caption{Concept proposed.
    In Raman microspectroscopy, the hyperspectrum tensor (A) has a matrix representation $\mathbf{H}$ (B, leftmost panel). $\mathbf{H}$ can be factorized into matrices $\mathbf{U}$, $\mathbf{S}$, $\mathbf{V}$, which have considerably lower sizes than $\mathbf{H}$, that is, $\mathbf{H}$ is low-rank. Therefore, one can considerably undersample $\mathbf{H}$ (C) and use recent signal processing algorithms so-called matrix completion for computational reconstruction. In our development (D), instead of using a spectrometer with costly cameras (left panel), $\mathbf{H}$ is under-sampled at high speed using a programmable spectrometer (right panel) that selects spectral bins (in the canonical representation), or a combination of spectral bins (in the multiplex approach), at random, but ensures uniform spatial sampling for high-sensitivity imaging. The lower panels illustrate the image plane spatial scanning and upper panels spectrometer wavelength sampling.
}
    \label{fig1}
\end{figure*}

Raman imaging is a simple label-free approach that exploits the intrinsic vibrational spectra of species as their fingerprint. It has been widely applied in various biological specimens~\cite{Shipp2017}, ranging from chemically selective imaging of cells~\cite{Matthaus2007,Klein2012,Okada2012,Muller2016} to \textit{spectroscopic} detection of pathologies~\cite{Harmsen2015,Jermyn2015}, bacteria~\cite{Bodelon2016,Lorenz2017}, and algae~\cite{Wu1998}, to cite a few examples. However, the spontaneous Raman process is a weak effect, therefore demanding costly and sensitive multi-pixel cameras with dispersive spectrometers. Such cameras limit dynamic applications where fast imaging is required, due to slow readout speed, and associated electronic noise. Apart from this technological bottleneck, the huge data sizes of hyperspectroscopy are an issue for real-life applications, due to data storage and display of large specimens hyperspectra (3D object as presented in Fig.~\ref{fig1}.A).

Recently, compressive Raman has been suggested to overcome data size and imaging speed limitations~\cite{Davis2011,Wilcox2012,Galvis-Carreno2014,Wei2016}. Compressive Raman is based on concepts of the emerging field of compressive sensing, which exploits new sampling paradigms based on experimental undersampling followed by computational reconstruction. In general, two strategies exist in compressive Raman: supervised~\cite{Wilcox2012,Refregier2018, Scotte2018,Corcoran2018} and unsupervised compression~\cite{Galvis-Carreno2014,Thompson2017}. Both concepts are based on the fact that the hyperspectrum $\mathbf{H}$ typically contains a small number of distinguishable chemical signatures, that is, it is extremely "chemically sparse" (Fig.~\ref{fig1}.B). Mathematically, this is equivalent to say that $\mathbf{H}$ is a low-rank matrix. Hence, from few data samples, the complete hyperspectrum can be reconstructed without loss of fidelity~\cite{Wilcox2012,Refregier2018}. On the one hand, the unsupervised approach is the most appealing as it requires low a priori information for computational reconstruction. However, current implementations of unsupervised methods are based on wide-field geometries~\cite{Thompson2017} and have unrealistic computational reconstruction times~\cite{August2013}; These two aspects preclude deep, dynamic and large-scale spectroscopic imaging of opaque biological samples. On the other hand, various demonstration of the supervised approach have shown high level of data compression with imaging speeds much faster than allowed by conventional camera-based technologies. The supervised method exploits the eigenspectra of $\mathbf{H}$ (Fig.~\ref{fig1}.B, rightmost spectra), as a priori information, to develop optimized spectral filters for fast, accurate, and precise chemical abundances determination~\cite{Wilcox2013,Refregier2018}. Nevertheless, this supervised method fails in chemically changing environments, as the "eigenspectra library" may evolve.

We present a new methodology based on the low-rank character of $\mathbf{H}$ to allow for fast chemical imaging of biological specimens. The method is based on a fast random undersampling scheme, which is a prerequisite for using the framework of matrix completion~\cite{Candes2009} (Fig.~\ref{fig1}.C). The key computational concept is based on the factorization $\mathbf{H}=\mathbf{U}\mathbf{S}\mathbf{V}^\intercal$, where $\mathbf{S}$ is a diagonal matrix related to the rank of $\mathbf{H}$, and the eigenvectors $\mathbf{U}$ and $\mathbf{V}$ represent the "eigenimages" and "eigenspectra", respectively (Fig.~\ref{fig1}.B). Since $\mathbf{H}$ is typically low-rank in Raman bio-imaging~\cite{Klein2012}, a completely sampled $\mathbf{H}$ means that highly redundant information is acquired; in other words, each spatial point is a simple linear combination of the few eigenspectra. Therefore, one can undersample $\mathbf{H}$ and use established and efficient algorithms of low-rank matrix completion estimation to "fill-in" the missing samples. A powerful advantage of the computational framework for chemical analysis is that it outputs a reduced dimension representation, which is ultimately desired for real-life applications (for instance, this could avoid a specialist in vibrational spectroscopy for interpretation).

\section*{Results}
We first describe the experimental methodology for fast random sampling scheme (Fig.~\ref{fig1}.D). Basically, it consists of a standard confocal Raman microscope, which allows for opaque samples observations, coupled to a recently developed high-throughput programmable spectrometer (see Methods)~\cite{Sturm2018}, thus enabling high sensitivity bio-imaging. Briefly, the costly cameras of a conventional spectrometer are replaced by a digital micromirror device (DMD) that can select wavelength bins to be detected with a highly sensitive single-pixel detector. As the focus moves, with step size smaller than the point-spread function (PSF) of the microscope, we concomitantly sample spectral bins randomly. Using this sampling strategy, we effectively sample several spectral components (alternatively, other basis can be used, \textit{e. g.} Hadamard basis) for every spatial pixel in the hyperspectral image. We then obtain a (raw) matrix that is transformed in the un-complete $\mathbf{H}_{exp}$, that is, in a spatial coordinates ($x,y$) vs. spectral coordinates ($\nu $) (see the missing boxes in Fig.~\ref{fig1}.C). By using this random hyperspectrum fast sampling methodology, $\mathbf{H}_{exp}$ can be readily processed using off-the-shelf algorithms~\cite{Candes2009,Xu2012}. The algorithm then outputs a singular value decomposition (SVD) of $\mathbf{H}$, which we use for post-processing in a conventional manner used in Raman imaging. Alternatively, we also use non-negative matrix completion algorithms~\cite{Xu2012} based on the factorization $\mathbf{H}=\mathbf{X}\mathbf{Y}$, where $\mathbf{X}$ and $\mathbf{Y}$ are non-negative matrices, motivated by the fact that Raman spectra are necessarily non-negative. Hence, this constraint may help to get a better reconstruction.

\begin{figure*}[hb]
    \centering
    \includegraphics[width=2\columnwidth]{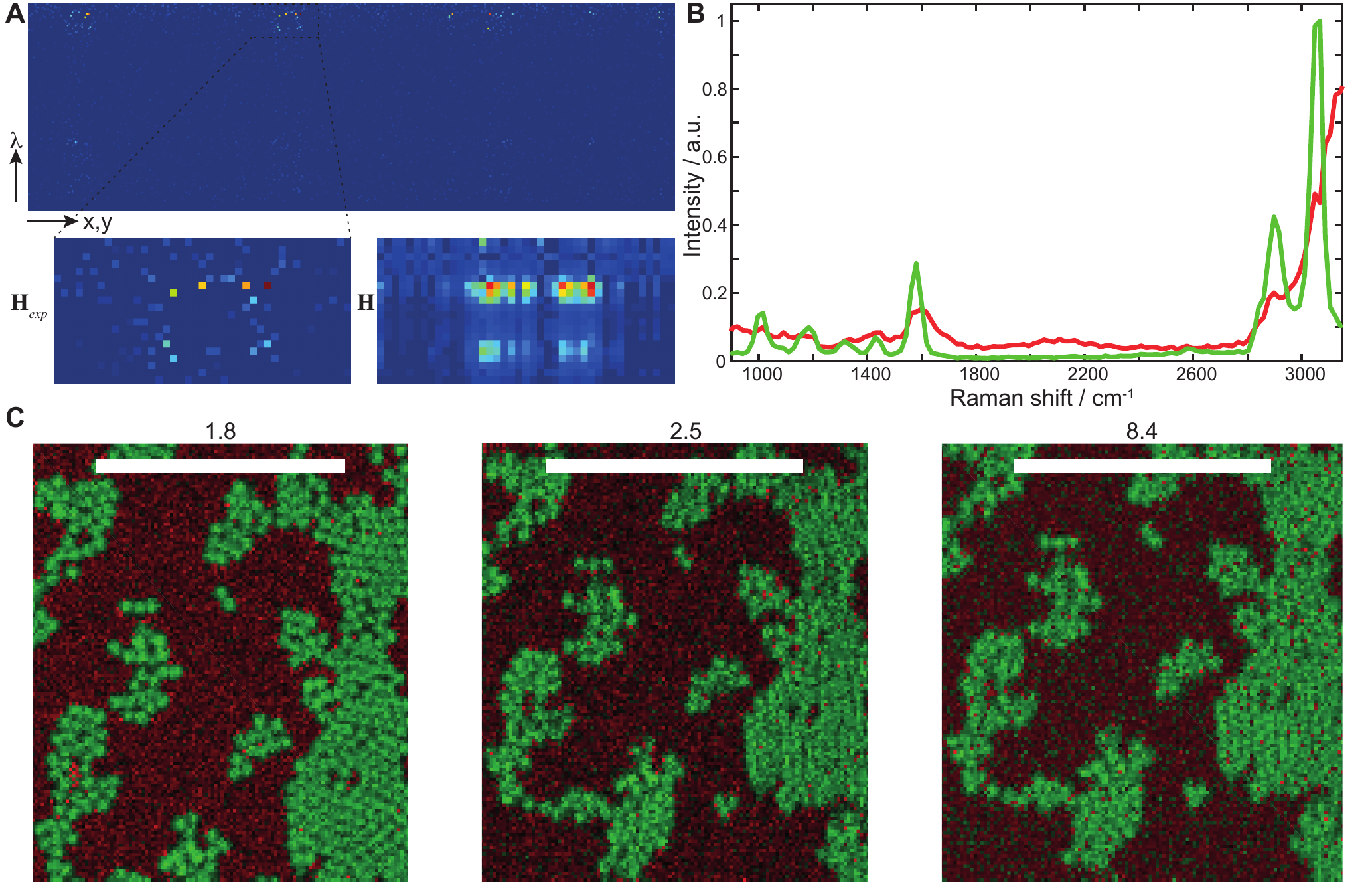}
    \caption{Proof-of-principle experiments.
    (A) Experimental hyperspectrum, before reconstruction, of polymer beads spread over a glass coverslip and embedded in water. In this particular example, only about 12\% of experimental samples have been acquired, as clearly seen in the zoomed in region (bottom left panel). After computational reconstruction a high-fidelity hyperspectrum is achieved (bottom right panel).
    (B) Average spectra of regions containing beads (green) and water plus glass (red).
    (C) Chemically selective images merged, following the colorcode of (B), for different level of compression (indicated on top of each panel). Effective pixel dwell time (left to right): 5.7, 6.2, 7.3~ms. Scale bars: 20~$\mu$m. 
}
    \label{fig2}
\end{figure*}

This methodology allows for fast imaging, with quasi instantaneous reconstruction speeds in low signal-level scenarios, typical of biological specimens. We first benchmark the approach with standard polymer beads (see Methods). Fig.~\ref{fig2}.A shows a representative undersampled hyperspectrum ($\mathbf{H}_{exp}$) together with its recovered completed spectrum ($\mathbf{H}$). The averaged spectra (Fig.~\ref{fig2}.B) of the merged images (Fig.~\ref{fig2}.C) shows the characteristic peaks of polystyrene (green), and the background (red) as a superposition of the water and glass coverslip spectra. The combined pseudo-color image (Fig.~\ref{fig2}.C) reveals that both species are anti-correlated as expected (the beads are larger than the microscope PSF). Fig.~\ref{fig2}.C also shows the effect of compression ratio. High level compression is obtained at the expense of increased noise in the reconstruction~\cite{Berto2017}. This is clearly seen by the loss of fidelity upon compression, nevertheless, high chemically selective images are achieved. Finally, the key advantage of the matrix completion algorithm is its reconstruction time which eventually allows for dynamic specimens imaging. Using standard laptop computers, we reconstruct the complete hyperspectrum at a rate of 8~ms/pixel (spatial).

\begin{figure*}[htb]
    \centering
    \includegraphics[width=2\columnwidth]{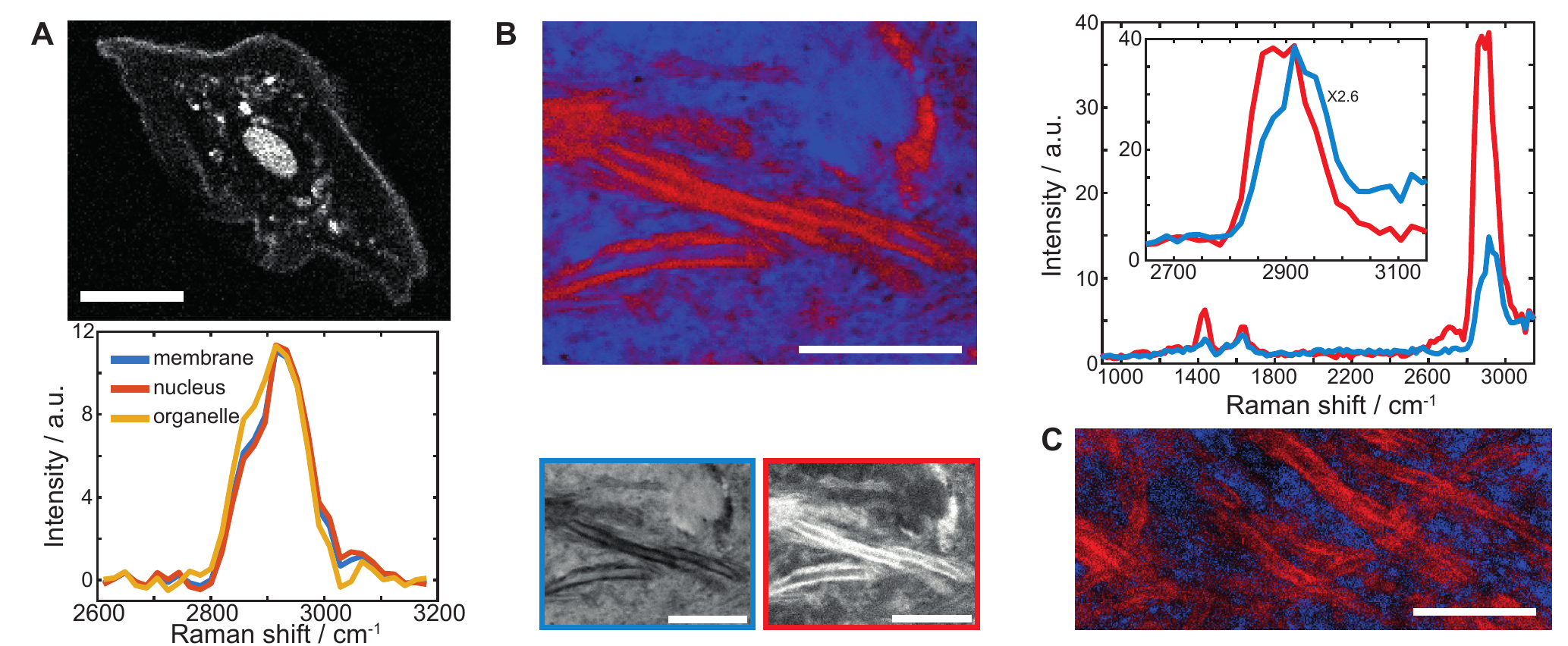}
    \caption{High-sensitivity bio-imaging of opaque and optically clear specimens.
    (A) High-sensitivity image of a cheek cell from integrated C-H stretch spectral region (2800-3000~cm$^{-1}$). (lower panel) The averaged background-corrected spectra in three regions of the cell, namely nucleus, membrane and round organelles. Effective pixel dwell time: 18.8~ms. Compression: 2.4.
    (B) Compressed Raman microspectroscopy of opaque brain tissue. (upper left panel) lipid-rich (red) and protein-rich (green) images, merged from their eigenimages (lower panels). The respective averaged spectra based on the eigenimages (upper right panel), highlighting the lipid-rich and protein-rich C-H stretch spectra differences (inset). Effective pixel dwell time: 19.6~ms. Compression: 2.4.
    (C) Supervised compressive Raman microspectroscopy images, using optimized filters based on the eigenspectra in (B). Note that samples in (B) and (C) are different, however both are from cerebellum tissues. Effective pixel dwell time: 0.4~ms. Compression: 64.
    Scale bars: 20~$\mu$m. 
}
    \label{fig3}
\end{figure*}

The confocal geometry used here allows high-sensitivity imaging of biological specimens with z-sectioning. We imaged cheek cells as a demonstration for cell sensitivity microspectroscopy (Fig.~\ref{fig3}.A). For that, we only spectrally scanned the C-H stretch region as previous studies have shown this spectral region to be sensitive for cell compartments analysis~\cite{Matthaus2007,Krafft2005}. The images generated from integrated C-H stretch peaks reveal multiple morphological features, such as the nucleus, membranes and small organelles. Closer inspection of average spectra of selected locations (Fig.~\ref{fig3}.A, lower panel) shows the expected trend in the C-H stretch intensity ratios $R=\frac{I_{2930}}{I_{2850}}$, which has been previously shown to report on the protein (high $R$) and lipid (low $R$) content. We observe that the organelle could be potentially assigned to lipid droplets and, interestingly, the cell membrane contains an intermediate ratio suggesting a mixture of proteins and lipid membranes. We also imaged opaque brain slices, recovering the expected sample morphology of tubular structures surrounded by continuous regions (Fig.~\ref{fig3}.B). The eigenimages from a SVD analysis (Fig.~\ref{fig3}.B, lower left panels) show that the most significant species are separated in tubular and non-tubular morphologies. Remarkably, the averaged spectra (Fig.~\ref{fig3}.B, top right panel) based on these images reveal that these two structures have high lipid content (green, low $R$) and high protein content (red, high $R$), in agreement with the chemical morphology of brain tissues: tubular structures are myelins made of lipids surrounding the axons rich in protein. 

\section*{Discussion}
We presented an unsupervised compressive Raman approach that enabled compressed imaging of biological specimens. The success relies in the combination of a new scanning methodology with the matrix completion framework. Since the method is based on a confocal geometry, it could in principle be used for imaging opaque samples at depth.  It does not require any a priori knowledge of the hyperspectrum, apart from being low-rank which, in fact, is fulfilled even under chemically complex scenario of biological specimens~\cite{Klein2012}. Furthermore, the computational framework provides fast reconstruction, suitable for imaging dynamic and large-scale specimens, two aspects that are often faced when imaging biological tissues. Such high-speed reconstruction could not be achieved in previous algorithms of unsupervised compressive spectroscopy as they require storage and multiplication of full rank matrices (\textit{i. e.} leading to slow reconstruction and large memory consumption)~\cite{August2013}. Finally, similar to previous compressive sensing algorithms, we observed that higher compression lowers reconstruction fidelity~\cite{Berto2017}, however, a highly chemically contrasted image could still be obtained. Note that spatial averaging allows to obtain cleaner spectra that can be used with the supervised approaches (see below).

Future modifications of the spatial scanning methodology can provide considerably higher speeds. In the current implementation, the effective pixel dwell time was limited by the scanning stage, rather than photon budget, meaning that ultimately the method can be shot-noise limited at high-speeds. Therefore, higher frame rates can be achieved by exploiting galvo scanners. We note that documented scanning methods of supervised approaches~\cite{Wilcox2013,Scotte2018,Corcoran2018} were based on a single spectral realization per image. Hence, they are not compatible with the matrix completion framework.

In conclusion, we have presented a new methodology for enabling compressive Raman bio-imaging. Apart from the inherent data size compression, the method is fast with reconstruction time negligible compared to image acquisition time, and inexpensive compared with alternative methods, with potential for much faster imaging speeds. We further showed that it is compatible with opaque samples imaging. We anticipate that the most powerful application will be to use the unsupervised approach as an input for the supervised ones, since it can lead to few seconds imaging speeds. This aspect is demonstrated in a proof-of-principle experiment presented in Fig.~\ref{fig3}.C: based on the eigenspectra learned from one tissue (Fig.~\ref{fig3}.B), we could image a second tissue at high-speed (Fig.~\ref{fig3}.C) exploiting optimized spectral filters (supervised compressive Raman)~\cite{Wilcox2012,Scotte2018}. This combination of methods resulted in 64 times data compression, and imaging speeds surpassing current camera-based technology. Therefore, the methodology presented here paves the way for fast clinical imaging using the inexpensive spontaneous Raman effect.


\section*{Methods}
\noindent \textbf{Sample preparation}. Polymer beads were prepared by drop-casting colloidal suspensions, with low polydispersity (Polysciences Inc.), on a coverslip sealed with water. Mouse cerebellum brain slices (thickness 500~$\mu$m) were fixed in an agarose solution with pbs buffer and sodium azide. Cheek cells were extracted from a male donor, and further dispersed with physiological saline solution to reduce the local concentration of debris before imaging. No fixation method was used for the cell imaging.

\noindent \textbf{Optical set-up}. A thourough description of the Raman microscope and its high-throughput compressive spectrometer can be found elsewhere~\cite{Donaldson2018,Sturm2018}. Briefly, a 532~nm-wavelength excitation laser (Oxxius LCX-532) is steered into an inverted microscope (Nikon Eclipse Ti-U) equipped with a high-NA oil-immersion objective (Nikon 60X/1.4NA). Samples are scanned with a nm-resolution piezoelectric translation-stage (Physik Instrumente P-545.3R7). The inelastically scattered light is guided with a multimode fiber into the home-built compressive spectrometer~\cite{Sturm2018}, based on a traditional Czerny-Turner design, however exchanging the usual high-sensitivity cameras for a programmable spectral filter (DMD, V-7001, ViALUX, 0.7" XGA resolution) which selects the various wavelength to be detected by a photon-counter module (SPCM-AQRH-44, Excelitas Technologies). The master clock is provided by the scanner stage, which triggers the exposition of the spectral masks (DMD) and detector acquisition.

\noindent \textbf{Computational techniques}. Different algorithms were used for the analysis. They are based on either soft-threshold SVD~\cite{Cai2010} or non-negative matrix factorization~\cite{Xu2012}. In general, we chose reconstructions with ranks between 2-5, as previous results have suggested~\cite{Klein2012}. In practice, we noticed that higher rank solutions only added noise to the reconstructed hyperspectrum. For Fig.~\ref{fig3}.C, the output of the matrix completion was passed to a standard SVD algorithm, to generate the eigenimages, in turn used for generating the spectra (Fig.~\ref{fig3}.B) for input of the supervised approach~\cite{Refregier2018,Scotte2018}. For the spectral sampling domain, we have used two spectral basis set for the spectral domain: a canonical (Fig.~\ref{fig2}) and Hadamard basis (Fig.~\ref{fig3}).

\section*{Acknowledgements} 
We thank Marie-Staphane Aigrot for kindly providing the brain slice samples. H.B.A. was supported by LabEX ENS-ICFP: ANR-10-LABX-0010/ANR-10-IDEX-0001-02 PSL*. This works was supported by European Research Council (ERC) (724473), Universitat Jaume I (PREDOC/2013/32), Generalitat Valenciana (PROMETEO/2016/079), Ministerio de Economia y Competitividad (MINECO, FIS2016-75618-R). S.G. is a member of the Institut Universitaire de France.\\

\section*{Author contributions}
All authors contributed to the writing and editing of the manuscript.

\section*{Competing financial interests}
The authors declare no competing financial interests.



\begin{thebibliography}{27}%
\makeatletter
\providecommand \@ifxundefined [1]{%
 \@ifx{#1\undefined}
}%
\providecommand \@ifnum [1]{%
 \ifnum #1\expandafter \@firstoftwo
 \else \expandafter \@secondoftwo
 \fi
}%
\providecommand \@ifx [1]{%
 \ifx #1\expandafter \@firstoftwo
 \else \expandafter \@secondoftwo
 \fi
}%
\providecommand \natexlab [1]{#1}%
\providecommand \enquote  [1]{``#1''}%
\providecommand \bibnamefont  [1]{#1}%
\providecommand \bibfnamefont [1]{#1}%
\providecommand \citenamefont [1]{#1}%
\providecommand \href@noop [0]{\@secondoftwo}%
\providecommand \href [0]{\begingroup \@sanitize@url \@href}%
\providecommand \@href[1]{\@@startlink{#1}\@@href}%
\providecommand \@@href[1]{\endgroup#1\@@endlink}%
\providecommand \@sanitize@url [0]{\catcode `\\12\catcode `\$12\catcode
  `\&12\catcode `\#12\catcode `\^12\catcode `\_12\catcode `\%12\relax}%
\providecommand \@@startlink[1]{}%
\providecommand \@@endlink[0]{}%
\providecommand \url  [0]{\begingroup\@sanitize@url \@url }%
\providecommand \@url [1]{\endgroup\@href {#1}{\urlprefix }}%
\providecommand \urlprefix  [0]{URL }%
\providecommand \Eprint [0]{\href }%
\providecommand \doibase [0]{http://dx.doi.org/}%
\providecommand \selectlanguage [0]{\@gobble}%
\providecommand \bibinfo  [0]{\@secondoftwo}%
\providecommand \bibfield  [0]{\@secondoftwo}%
\providecommand \translation [1]{[#1]}%
\providecommand \BibitemOpen [0]{}%
\providecommand \bibitemStop [0]{}%
\providecommand \bibitemNoStop [0]{.\EOS\space}%
\providecommand \EOS [0]{\spacefactor3000\relax}%
\providecommand \BibitemShut  [1]{\csname bibitem#1\endcsname}%
\let\auto@bib@innerbib\@empty
\bibitem [{\citenamefont {Shipp}\ \emph {et~al.}(2017)\citenamefont {Shipp},
  \citenamefont {Sinjab},\ and\ \citenamefont {Notingher}}]{Shipp2017}%
  \BibitemOpen
  \bibfield  {author} {\bibinfo {author} {\bibfnamefont {Dustin~W}\
  \bibnamefont {Shipp}}, \bibinfo {author} {\bibfnamefont {Faris}\ \bibnamefont
  {Sinjab}}, \ and\ \bibinfo {author} {\bibfnamefont {Ioan}\ \bibnamefont
  {Notingher}},\ }\bibfield  {title} {\enquote {\bibinfo {title} {Raman
  spectroscopy: Techniques and applications in the life sciences},}\
  }\href@noop {} {\bibfield  {journal} {\bibinfo  {journal} {Advances in Optics
  and Photonics}\ }\textbf {\bibinfo {volume} {9}},\ \bibinfo {pages}
  {315--428} (\bibinfo {year} {2017})}\BibitemShut {NoStop}%
\bibitem [{\citenamefont {Matth{\"a}us}\ \emph {et~al.}(2007)\citenamefont
  {Matth{\"a}us}, \citenamefont {Chernenko}, \citenamefont {Newmark},
  \citenamefont {Warner},\ and\ \citenamefont {Diem}}]{Matthaus2007}%
  \BibitemOpen
  \bibfield  {author} {\bibinfo {author} {\bibfnamefont {Christian}\
  \bibnamefont {Matth{\"a}us}}, \bibinfo {author} {\bibfnamefont {Tatyana}\
  \bibnamefont {Chernenko}}, \bibinfo {author} {\bibfnamefont {Judith~A}\
  \bibnamefont {Newmark}}, \bibinfo {author} {\bibfnamefont {Carol~M}\
  \bibnamefont {Warner}}, \ and\ \bibinfo {author} {\bibfnamefont {Max}\
  \bibnamefont {Diem}},\ }\bibfield  {title} {\enquote {\bibinfo {title}
  {Label-free detection of mitochondrial distribution in cells by nonresonant
  raman microspectroscopy},}\ }\href@noop {} {\bibfield  {journal} {\bibinfo
  {journal} {Biophysical Journal}\ }\textbf {\bibinfo {volume} {93}},\ \bibinfo
  {pages} {668--673} (\bibinfo {year} {2007})}\BibitemShut {NoStop}%
\bibitem [{\citenamefont {Klein}\ \emph {et~al.}(2012)\citenamefont {Klein},
  \citenamefont {Gigler}, \citenamefont {Aschenbrenner}, \citenamefont
  {Monetti}, \citenamefont {Bunk}, \citenamefont {Jamitzky}, \citenamefont
  {Morfill}, \citenamefont {Stark},\ and\ \citenamefont
  {Schlegel}}]{Klein2012}%
  \BibitemOpen
  \bibfield  {author} {\bibinfo {author} {\bibfnamefont {Katharina}\
  \bibnamefont {Klein}}, \bibinfo {author} {\bibfnamefont {Alexander~M}\
  \bibnamefont {Gigler}}, \bibinfo {author} {\bibfnamefont {Thomas}\
  \bibnamefont {Aschenbrenner}}, \bibinfo {author} {\bibfnamefont {Roberto}\
  \bibnamefont {Monetti}}, \bibinfo {author} {\bibfnamefont {Wolfram}\
  \bibnamefont {Bunk}}, \bibinfo {author} {\bibfnamefont {Ferdinand}\
  \bibnamefont {Jamitzky}}, \bibinfo {author} {\bibfnamefont {Gregor}\
  \bibnamefont {Morfill}}, \bibinfo {author} {\bibfnamefont {Robert~W}\
  \bibnamefont {Stark}}, \ and\ \bibinfo {author} {\bibfnamefont {J{\"u}rgen}\
  \bibnamefont {Schlegel}},\ }\bibfield  {title} {\enquote {\bibinfo {title}
  {Label-free live-cell imaging with confocal raman microscopy},}\ }\href@noop
  {} {\bibfield  {journal} {\bibinfo  {journal} {Biophysical Journal}\ }\textbf
  {\bibinfo {volume} {102}},\ \bibinfo {pages} {360--368} (\bibinfo {year}
  {2012})}\BibitemShut {NoStop}%
\bibitem [{\citenamefont {Okada}\ \emph {et~al.}(2012)\citenamefont {Okada},
  \citenamefont {Smith}, \citenamefont {Palonpon}, \citenamefont {Endo},
  \citenamefont {Kawata}, \citenamefont {Sodeoka},\ and\ \citenamefont
  {Fujita}}]{Okada2012}%
  \BibitemOpen
  \bibfield  {author} {\bibinfo {author} {\bibfnamefont {Masaya}\ \bibnamefont
  {Okada}}, \bibinfo {author} {\bibfnamefont {Nicholas~Isaac}\ \bibnamefont
  {Smith}}, \bibinfo {author} {\bibfnamefont {Almar~Flotildes}\ \bibnamefont
  {Palonpon}}, \bibinfo {author} {\bibfnamefont {Hiromi}\ \bibnamefont {Endo}},
  \bibinfo {author} {\bibfnamefont {Satoshi}\ \bibnamefont {Kawata}}, \bibinfo
  {author} {\bibfnamefont {Mikiko}\ \bibnamefont {Sodeoka}}, \ and\ \bibinfo
  {author} {\bibfnamefont {Katsumasa}\ \bibnamefont {Fujita}},\ }\bibfield
  {title} {\enquote {\bibinfo {title} {Label-free raman observation of
  cytochrome c dynamics during apoptosis},}\ }\href@noop {} {\bibfield
  {journal} {\bibinfo  {journal} {Proc. Natl. Acad. Sci. U. S. A.}\ }\textbf
  {\bibinfo {volume} {109}},\ \bibinfo {pages} {28--32} (\bibinfo {year}
  {2012})}\BibitemShut {NoStop}%
\bibitem [{\citenamefont {M\"{u}ller}\ \emph {et~al.}(2016)\citenamefont
  {M\"{u}ller}, \citenamefont {Kielhorn}, \citenamefont {Schmitt},
  \citenamefont {Popp},\ and\ \citenamefont {Heintzmann}}]{Muller2016}%
  \BibitemOpen
  \bibfield  {author} {\bibinfo {author} {\bibfnamefont {Walter}\ \bibnamefont
  {M\"{u}ller}}, \bibinfo {author} {\bibfnamefont {Martin}\ \bibnamefont
  {Kielhorn}}, \bibinfo {author} {\bibfnamefont {Michael}\ \bibnamefont
  {Schmitt}}, \bibinfo {author} {\bibfnamefont {J\"{u}rgen}\ \bibnamefont
  {Popp}}, \ and\ \bibinfo {author} {\bibfnamefont {Rainer}\ \bibnamefont
  {Heintzmann}},\ }\bibfield  {title} {\enquote {\bibinfo {title} {Light sheet
  raman micro-spectroscopy},}\ }\href {\doibase 10.1364/OPTICA.3.000452}
  {\bibfield  {journal} {\bibinfo  {journal} {Optica}\ }\textbf {\bibinfo
  {volume} {3}},\ \bibinfo {pages} {452--457} (\bibinfo {year}
  {2016})}\BibitemShut {NoStop}%
\bibitem [{\citenamefont {Harmsen}\ \emph {et~al.}(2015)\citenamefont
  {Harmsen}, \citenamefont {Huang}, \citenamefont {Wall}, \citenamefont
  {Karabeber}, \citenamefont {Samii}, \citenamefont {Spaliviero}, \citenamefont
  {White}, \citenamefont {Monette}, \citenamefont {O{\textquoteright}Connor},
  \citenamefont {Pitter}, \citenamefont {Sastra}, \citenamefont {Saborowski},
  \citenamefont {Holland}, \citenamefont {Singer}, \citenamefont {Olive},
  \citenamefont {Lowe}, \citenamefont {Blasberg},\ and\ \citenamefont
  {Kircher}}]{Harmsen2015}%
  \BibitemOpen
  \bibfield  {author} {\bibinfo {author} {\bibfnamefont {Stefan}\ \bibnamefont
  {Harmsen}}, \bibinfo {author} {\bibfnamefont {Ruimin}\ \bibnamefont {Huang}},
  \bibinfo {author} {\bibfnamefont {Matthew~A.}\ \bibnamefont {Wall}}, \bibinfo
  {author} {\bibfnamefont {Hazem}\ \bibnamefont {Karabeber}}, \bibinfo {author}
  {\bibfnamefont {Jason~M.}\ \bibnamefont {Samii}}, \bibinfo {author}
  {\bibfnamefont {Massimiliano}\ \bibnamefont {Spaliviero}}, \bibinfo {author}
  {\bibfnamefont {Julie~R.}\ \bibnamefont {White}}, \bibinfo {author}
  {\bibfnamefont {S{\'e}bastien}\ \bibnamefont {Monette}}, \bibinfo {author}
  {\bibfnamefont {Rachael}\ \bibnamefont {O{\textquoteright}Connor}}, \bibinfo
  {author} {\bibfnamefont {Kenneth~L.}\ \bibnamefont {Pitter}}, \bibinfo
  {author} {\bibfnamefont {Stephen~A.}\ \bibnamefont {Sastra}}, \bibinfo
  {author} {\bibfnamefont {Michael}\ \bibnamefont {Saborowski}}, \bibinfo
  {author} {\bibfnamefont {Eric~C.}\ \bibnamefont {Holland}}, \bibinfo {author}
  {\bibfnamefont {Samuel}\ \bibnamefont {Singer}}, \bibinfo {author}
  {\bibfnamefont {Kenneth~P.}\ \bibnamefont {Olive}}, \bibinfo {author}
  {\bibfnamefont {Scott~W.}\ \bibnamefont {Lowe}}, \bibinfo {author}
  {\bibfnamefont {Ronald~G.}\ \bibnamefont {Blasberg}}, \ and\ \bibinfo
  {author} {\bibfnamefont {Moritz~F.}\ \bibnamefont {Kircher}},\ }\bibfield
  {title} {\enquote {\bibinfo {title} {Surface-enhanced resonance raman
  scattering nanostars for high-precision cancer imaging},}\ }\href {\doibase
  10.1126/scitranslmed.3010633} {\bibfield  {journal} {\bibinfo  {journal}
  {Science Translational Medicine}\ }\textbf {\bibinfo {volume} {7}},\ \bibinfo
  {pages} {271ra7} (\bibinfo {year} {2015})},\ \Eprint
  {http://arxiv.org/abs/http://stm.sciencemag.org/content/7/271/271ra7.full.pdf}
  {http://stm.sciencemag.org/content/7/271/271ra7.full.pdf} \BibitemShut
  {NoStop}%
\bibitem [{\citenamefont {Jermyn}\ \emph {et~al.}(2015)\citenamefont {Jermyn},
  \citenamefont {Mok}, \citenamefont {Mercier}, \citenamefont {Desroches},
  \citenamefont {Pichette}, \citenamefont {Saint-Arnaud}, \citenamefont
  {Bernstein}, \citenamefont {Guiot}, \citenamefont {Petrecca},\ and\
  \citenamefont {Leblond}}]{Jermyn2015}%
  \BibitemOpen
  \bibfield  {author} {\bibinfo {author} {\bibfnamefont {Michael}\ \bibnamefont
  {Jermyn}}, \bibinfo {author} {\bibfnamefont {Kelvin}\ \bibnamefont {Mok}},
  \bibinfo {author} {\bibfnamefont {Jeanne}\ \bibnamefont {Mercier}}, \bibinfo
  {author} {\bibfnamefont {Joannie}\ \bibnamefont {Desroches}}, \bibinfo
  {author} {\bibfnamefont {Julien}\ \bibnamefont {Pichette}}, \bibinfo {author}
  {\bibfnamefont {Karl}\ \bibnamefont {Saint-Arnaud}}, \bibinfo {author}
  {\bibfnamefont {Liane}\ \bibnamefont {Bernstein}}, \bibinfo {author}
  {\bibfnamefont {Marie-Christine}\ \bibnamefont {Guiot}}, \bibinfo {author}
  {\bibfnamefont {Kevin}\ \bibnamefont {Petrecca}}, \ and\ \bibinfo {author}
  {\bibfnamefont {Frederic}\ \bibnamefont {Leblond}},\ }\bibfield  {title}
  {\enquote {\bibinfo {title} {Intraoperative brain cancer detection with raman
  spectroscopy in humans},}\ }\href {\doibase 10.1126/scitranslmed.aaa2384}
  {\bibfield  {journal} {\bibinfo  {journal} {Science Translational Medicine}\
  }\textbf {\bibinfo {volume} {7}},\ \bibinfo {pages} {274ra19} (\bibinfo
  {year} {2015})}\BibitemShut {NoStop}%
\bibitem [{\citenamefont {Bodel{\'o}n}\ \emph {et~al.}(2016)\citenamefont
  {Bodel{\'o}n}, \citenamefont {Montes-Garc{\'\i}a}, \citenamefont
  {L{\'o}pez-Puente}, \citenamefont {Hill}, \citenamefont {Hamon},
  \citenamefont {Sanz-Ortiz}, \citenamefont {Rodal-Cedeira}, \citenamefont
  {Costas}, \citenamefont {Celiksoy}, \citenamefont {P{\'e}rez-Juste} \emph
  {et~al.}}]{Bodelon2016}%
  \BibitemOpen
  \bibfield  {author} {\bibinfo {author} {\bibfnamefont {Gustavo}\ \bibnamefont
  {Bodel{\'o}n}}, \bibinfo {author} {\bibfnamefont {Ver{\'o}nica}\ \bibnamefont
  {Montes-Garc{\'\i}a}}, \bibinfo {author} {\bibfnamefont {Vanesa}\
  \bibnamefont {L{\'o}pez-Puente}}, \bibinfo {author} {\bibfnamefont {Eric~H}\
  \bibnamefont {Hill}}, \bibinfo {author} {\bibfnamefont {Cyrille}\
  \bibnamefont {Hamon}}, \bibinfo {author} {\bibfnamefont {Marta~N}\
  \bibnamefont {Sanz-Ortiz}}, \bibinfo {author} {\bibfnamefont {Sergio}\
  \bibnamefont {Rodal-Cedeira}}, \bibinfo {author} {\bibfnamefont {Celina}\
  \bibnamefont {Costas}}, \bibinfo {author} {\bibfnamefont {Sirin}\
  \bibnamefont {Celiksoy}}, \bibinfo {author} {\bibfnamefont {Ignacio}\
  \bibnamefont {P{\'e}rez-Juste}},  \emph {et~al.},\ }\bibfield  {title}
  {\enquote {\bibinfo {title} {Detection and imaging of quorum sensing in
  pseudomonas aeruginosa biofilm communities by surface-enhanced resonance
  raman scattering},}\ }\href@noop {} {\bibfield  {journal} {\bibinfo
  {journal} {Nature Materials}\ }\textbf {\bibinfo {volume} {15}},\ \bibinfo
  {pages} {1203} (\bibinfo {year} {2016})}\BibitemShut {NoStop}%
\bibitem [{\citenamefont {Lorenz}\ \emph {et~al.}(2017)\citenamefont {Lorenz},
  \citenamefont {Wichmann}, \citenamefont {St{\"o}ckel}, \citenamefont
  {R{\"o}sch},\ and\ \citenamefont {Popp}}]{Lorenz2017}%
  \BibitemOpen
  \bibfield  {author} {\bibinfo {author} {\bibfnamefont {Bj{\"o}rn}\
  \bibnamefont {Lorenz}}, \bibinfo {author} {\bibfnamefont {Christina}\
  \bibnamefont {Wichmann}}, \bibinfo {author} {\bibfnamefont {Stephan}\
  \bibnamefont {St{\"o}ckel}}, \bibinfo {author} {\bibfnamefont {Petra}\
  \bibnamefont {R{\"o}sch}}, \ and\ \bibinfo {author} {\bibfnamefont
  {J{\"u}rgen}\ \bibnamefont {Popp}},\ }\bibfield  {title} {\enquote {\bibinfo
  {title} {Cultivation-free raman spectroscopic investigations of bacteria},}\
  }\href@noop {} {\bibfield  {journal} {\bibinfo  {journal} {Trends in
  Microbiology}\ }\textbf {\bibinfo {volume} {25}},\ \bibinfo {pages}
  {413--424} (\bibinfo {year} {2017})}\BibitemShut {NoStop}%
\bibitem [{\citenamefont {Wu}\ \emph {et~al.}(1998)\citenamefont {Wu},
  \citenamefont {Nelson}, \citenamefont {Hargraves}, \citenamefont {Zhang},
  \citenamefont {Brown},\ and\ \citenamefont {Seelenbinder}}]{Wu1998}%
  \BibitemOpen
  \bibfield  {author} {\bibinfo {author} {\bibfnamefont {Q}~\bibnamefont {Wu}},
  \bibinfo {author} {\bibfnamefont {WH}~\bibnamefont {Nelson}}, \bibinfo
  {author} {\bibfnamefont {P}~\bibnamefont {Hargraves}}, \bibinfo {author}
  {\bibfnamefont {J}~\bibnamefont {Zhang}}, \bibinfo {author} {\bibfnamefont
  {CW}~\bibnamefont {Brown}}, \ and\ \bibinfo {author} {\bibfnamefont
  {JA}~\bibnamefont {Seelenbinder}},\ }\bibfield  {title} {\enquote {\bibinfo
  {title} {Differentiation of algae clones on the basis of resonance raman
  spectra excited by visible light},}\ }\href@noop {} {\bibfield  {journal}
  {\bibinfo  {journal} {Analytical Chemistry}\ }\textbf {\bibinfo {volume}
  {70}},\ \bibinfo {pages} {1782--1787} (\bibinfo {year} {1998})}\BibitemShut
  {NoStop}%
\bibitem [{\citenamefont {Davis}\ \emph {et~al.}(2011)\citenamefont {Davis},
  \citenamefont {Hemphill}, \citenamefont {Cebeci~Maltas}, \citenamefont
  {Zipper}, \citenamefont {Wang},\ and\ \citenamefont {Ben-Amotz}}]{Davis2011}%
  \BibitemOpen
  \bibfield  {author} {\bibinfo {author} {\bibfnamefont {Brandon~M}\
  \bibnamefont {Davis}}, \bibinfo {author} {\bibfnamefont {Amanda~J}\
  \bibnamefont {Hemphill}}, \bibinfo {author} {\bibfnamefont {Derya}\
  \bibnamefont {Cebeci~Maltas}}, \bibinfo {author} {\bibfnamefont {Michael~A}\
  \bibnamefont {Zipper}}, \bibinfo {author} {\bibfnamefont {Ping}\ \bibnamefont
  {Wang}}, \ and\ \bibinfo {author} {\bibfnamefont {Dor}\ \bibnamefont
  {Ben-Amotz}},\ }\bibfield  {title} {\enquote {\bibinfo {title} {Multivariate
  hyperspectral raman imaging using compressive detection},}\ }\href@noop {}
  {\bibfield  {journal} {\bibinfo  {journal} {Analytical Chemistry}\ }\textbf
  {\bibinfo {volume} {83}},\ \bibinfo {pages} {5086--5092} (\bibinfo {year}
  {2011})}\BibitemShut {NoStop}%
\bibitem [{\citenamefont {Wilcox}\ \emph {et~al.}(2012)\citenamefont {Wilcox},
  \citenamefont {Buzzard}, \citenamefont {Lucier}, \citenamefont {Wang},\ and\
  \citenamefont {Ben-Amotz}}]{Wilcox2012}%
  \BibitemOpen
  \bibfield  {author} {\bibinfo {author} {\bibfnamefont {David~S}\ \bibnamefont
  {Wilcox}}, \bibinfo {author} {\bibfnamefont {Gregery~T}\ \bibnamefont
  {Buzzard}}, \bibinfo {author} {\bibfnamefont {Bradley~J}\ \bibnamefont
  {Lucier}}, \bibinfo {author} {\bibfnamefont {Ping}\ \bibnamefont {Wang}}, \
  and\ \bibinfo {author} {\bibfnamefont {Dor}\ \bibnamefont {Ben-Amotz}},\
  }\bibfield  {title} {\enquote {\bibinfo {title} {Photon level chemical
  classification using digital compressive detection},}\ }\href@noop {}
  {\bibfield  {journal} {\bibinfo  {journal} {Analytica Chimica Acta}\ }\textbf
  {\bibinfo {volume} {755}},\ \bibinfo {pages} {17--27} (\bibinfo {year}
  {2012})}\BibitemShut {NoStop}%
\bibitem [{\citenamefont {Galvis-Carre{\~n}o}\ \emph
  {et~al.}(2014)\citenamefont {Galvis-Carre{\~n}o}, \citenamefont
  {Mej{\'\i}a-Melgarejo},\ and\ \citenamefont
  {Arguello-Fuentes}}]{Galvis-Carreno2014}%
  \BibitemOpen
  \bibfield  {author} {\bibinfo {author} {\bibfnamefont {Diana~Fernanda}\
  \bibnamefont {Galvis-Carre{\~n}o}}, \bibinfo {author} {\bibfnamefont
  {Yuri~Hercilia}\ \bibnamefont {Mej{\'\i}a-Melgarejo}}, \ and\ \bibinfo
  {author} {\bibfnamefont {Henry}\ \bibnamefont {Arguello-Fuentes}},\
  }\bibfield  {title} {\enquote {\bibinfo {title} {Efficient reconstruction of
  raman spectroscopy imaging based on compressive sensing},}\ }\href@noop {}
  {\bibfield  {journal} {\bibinfo  {journal} {Dyna}\ }\textbf {\bibinfo
  {volume} {81}},\ \bibinfo {pages} {116--124} (\bibinfo {year}
  {2014})}\BibitemShut {NoStop}%
\bibitem [{\citenamefont {Wei}\ \emph {et~al.}(2016)\citenamefont {Wei},
  \citenamefont {Chen}, \citenamefont {Ong}, \citenamefont {Perlaki},\ and\
  \citenamefont {Liu}}]{Wei2016}%
  \BibitemOpen
  \bibfield  {author} {\bibinfo {author} {\bibfnamefont {Dong}\ \bibnamefont
  {Wei}}, \bibinfo {author} {\bibfnamefont {Shuo}\ \bibnamefont {Chen}},
  \bibinfo {author} {\bibfnamefont {Yi~Hong}\ \bibnamefont {Ong}}, \bibinfo
  {author} {\bibfnamefont {Clint}\ \bibnamefont {Perlaki}}, \ and\ \bibinfo
  {author} {\bibfnamefont {Quan}\ \bibnamefont {Liu}},\ }\bibfield  {title}
  {\enquote {\bibinfo {title} {Fast wide-field raman spectroscopic imaging
  based on simultaneous multi-channel image acquisition and wiener
  estimation},}\ }\href@noop {} {\bibfield  {journal} {\bibinfo  {journal}
  {Optics Letters}\ }\textbf {\bibinfo {volume} {41}},\ \bibinfo {pages}
  {2783--2786} (\bibinfo {year} {2016})}\BibitemShut {NoStop}%
\bibitem [{\citenamefont {R{\'e}fr{\'e}gier}\ \emph {et~al.}(2018)\citenamefont
  {R{\'e}fr{\'e}gier}, \citenamefont {Scott\'{e}}, \citenamefont {de~Aguiar},
  \citenamefont {Rigneault},\ and\ \citenamefont {Galland}}]{Refregier2018}%
  \BibitemOpen
  \bibfield  {author} {\bibinfo {author} {\bibfnamefont {P.}~\bibnamefont
  {R{\'e}fr{\'e}gier}}, \bibinfo {author} {\bibfnamefont {C.}~\bibnamefont
  {Scott\'{e}}}, \bibinfo {author} {\bibfnamefont {H.~B.}\ \bibnamefont
  {de~Aguiar}}, \bibinfo {author} {\bibfnamefont {H.}~\bibnamefont
  {Rigneault}}, \ and\ \bibinfo {author} {\bibfnamefont {F.}~\bibnamefont
  {Galland}},\ }\bibfield  {title} {\enquote {\bibinfo {title} {Precision of
  proportion estimation with binary compressed raman spectrum},}\ }\href@noop
  {} {\bibfield  {journal} {\bibinfo  {journal} {J. Opt. Soc. Am. A}\ }\textbf
  {\bibinfo {volume} {35}},\ \bibinfo {pages} {125--134} (\bibinfo {year}
  {2018})}\BibitemShut {NoStop}%
\bibitem [{\citenamefont {Scott\'{e}}\ \emph {et~al.}(2018)\citenamefont
  {Scott\'{e}}, \citenamefont {de~Aguiar}, \citenamefont {Marguet},
  \citenamefont {Green}, \citenamefont {Bouzy}, \citenamefont {Vergnole},
  \citenamefont {Winlove}, \citenamefont {Stone},\ and\ \citenamefont
  {Rigneault}}]{Scotte2018}%
  \BibitemOpen
  \bibfield  {author} {\bibinfo {author} {\bibfnamefont {C.}~\bibnamefont
  {Scott\'{e}}}, \bibinfo {author} {\bibfnamefont {H.~B.}\ \bibnamefont
  {de~Aguiar}}, \bibinfo {author} {\bibfnamefont {D.}~\bibnamefont {Marguet}},
  \bibinfo {author} {\bibfnamefont {E.}~\bibnamefont {Green}}, \bibinfo
  {author} {\bibfnamefont {P.}~\bibnamefont {Bouzy}}, \bibinfo {author}
  {\bibfnamefont {S.}~\bibnamefont {Vergnole}}, \bibinfo {author}
  {\bibfnamefont {P.}~\bibnamefont {Winlove}}, \bibinfo {author} {\bibfnamefont
  {N.}~\bibnamefont {Stone}}, \ and\ \bibinfo {author} {\bibfnamefont
  {H.}~\bibnamefont {Rigneault}},\ }\bibfield  {title} {\enquote {\bibinfo
  {title} {Assessment of compressive raman versus hyperspectral raman for
  microcalcification chemical imaging},}\ }\href@noop {} {\bibfield  {journal}
  {\bibinfo  {journal} {Anal. Chem.}\ }\textbf {\bibinfo {volume} {90}},\
  \bibinfo {pages} {7197--7203} (\bibinfo {year} {2018})}\BibitemShut {NoStop}%
\bibitem [{\citenamefont {Corcoran}(2018)}]{Corcoran2018}%
  \BibitemOpen
  \bibfield  {author} {\bibinfo {author} {\bibfnamefont {Timothy~C}\
  \bibnamefont {Corcoran}},\ }\bibfield  {title} {\enquote {\bibinfo {title}
  {Compressive detection of highly overlapped spectra using
  walsh--hadamard-based filter functions},}\ }\href@noop {} {\bibfield
  {journal} {\bibinfo  {journal} {Applied Spectroscopy}\ }\textbf {\bibinfo
  {volume} {72}},\ \bibinfo {pages} {392--403} (\bibinfo {year}
  {2018})}\BibitemShut {NoStop}%
\bibitem [{\citenamefont {Thompson}\ \emph {et~al.}(2017)\citenamefont
  {Thompson}, \citenamefont {Bixler}, \citenamefont {Hokr}, \citenamefont
  {Noojin}, \citenamefont {Scully},\ and\ \citenamefont
  {Yakovlev}}]{Thompson2017}%
  \BibitemOpen
  \bibfield  {author} {\bibinfo {author} {\bibfnamefont {Jonathan~V}\
  \bibnamefont {Thompson}}, \bibinfo {author} {\bibfnamefont {Joel~N}\
  \bibnamefont {Bixler}}, \bibinfo {author} {\bibfnamefont {Brett~H}\
  \bibnamefont {Hokr}}, \bibinfo {author} {\bibfnamefont {Gary~D}\ \bibnamefont
  {Noojin}}, \bibinfo {author} {\bibfnamefont {Marlan~O}\ \bibnamefont
  {Scully}}, \ and\ \bibinfo {author} {\bibfnamefont {Vladislav~V}\
  \bibnamefont {Yakovlev}},\ }\bibfield  {title} {\enquote {\bibinfo {title}
  {Single-shot chemical detection and identification with compressed
  hyperspectral raman imaging},}\ }\href@noop {} {\bibfield  {journal}
  {\bibinfo  {journal} {Optics Letters}\ }\textbf {\bibinfo {volume} {42}},\
  \bibinfo {pages} {2169--2172} (\bibinfo {year} {2017})}\BibitemShut {NoStop}%
\bibitem [{\citenamefont {August}\ \emph {et~al.}(2013)\citenamefont {August},
  \citenamefont {Vachman}, \citenamefont {Rivenson},\ and\ \citenamefont
  {Stern}}]{August2013}%
  \BibitemOpen
  \bibfield  {author} {\bibinfo {author} {\bibfnamefont {Yitzhak}\ \bibnamefont
  {August}}, \bibinfo {author} {\bibfnamefont {Chaim}\ \bibnamefont {Vachman}},
  \bibinfo {author} {\bibfnamefont {Yair}\ \bibnamefont {Rivenson}}, \ and\
  \bibinfo {author} {\bibfnamefont {Adrian}\ \bibnamefont {Stern}},\ }\bibfield
   {title} {\enquote {\bibinfo {title} {Compressive hyperspectral imaging by
  random separable projections in both the spatial and the spectral domains},}\
  }\href@noop {} {\bibfield  {journal} {\bibinfo  {journal} {Applied Optics}\
  }\textbf {\bibinfo {volume} {52}},\ \bibinfo {pages} {D46--D54} (\bibinfo
  {year} {2013})}\BibitemShut {NoStop}%
\bibitem [{\citenamefont {Wilcox}\ \emph {et~al.}(2013)\citenamefont {Wilcox},
  \citenamefont {Buzzard}, \citenamefont {Lucier}, \citenamefont {Rehrauer},
  \citenamefont {Wang},\ and\ \citenamefont {Ben-Amotz}}]{Wilcox2013}%
  \BibitemOpen
  \bibfield  {author} {\bibinfo {author} {\bibfnamefont {David~S.}\
  \bibnamefont {Wilcox}}, \bibinfo {author} {\bibfnamefont {Gregery~T.}\
  \bibnamefont {Buzzard}}, \bibinfo {author} {\bibfnamefont {Bradley~J.}\
  \bibnamefont {Lucier}}, \bibinfo {author} {\bibfnamefont {Owen~G.}\
  \bibnamefont {Rehrauer}}, \bibinfo {author} {\bibfnamefont {Ping}\
  \bibnamefont {Wang}}, \ and\ \bibinfo {author} {\bibfnamefont {Dor}\
  \bibnamefont {Ben-Amotz}},\ }\bibfield  {title} {\enquote {\bibinfo {title}
  {Digital compressive chemical quantitation and hyperspectral imaging},}\
  }\href {\doibase 10.1039/C3AN00309D} {\bibfield  {journal} {\bibinfo
  {journal} {Analyst}\ }\textbf {\bibinfo {volume} {138}},\ \bibinfo {pages}
  {4982--4990} (\bibinfo {year} {2013})}\BibitemShut {NoStop}%
\bibitem [{\citenamefont {Cand{\`e}s}\ and\ \citenamefont
  {Recht}(2009)}]{Candes2009}%
  \BibitemOpen
  \bibfield  {author} {\bibinfo {author} {\bibfnamefont {Emmanuel~J}\
  \bibnamefont {Cand{\`e}s}}\ and\ \bibinfo {author} {\bibfnamefont {Benjamin}\
  \bibnamefont {Recht}},\ }\bibfield  {title} {\enquote {\bibinfo {title}
  {Exact matrix completion via convex optimization},}\ }\href@noop {}
  {\bibfield  {journal} {\bibinfo  {journal} {Foundations of Computational
  Mathematics}\ }\textbf {\bibinfo {volume} {9}},\ \bibinfo {pages} {717}
  (\bibinfo {year} {2009})}\BibitemShut {NoStop}%
\bibitem [{\citenamefont {Sturm}\ \emph {et~al.}(2018)\citenamefont {Sturm},
  \citenamefont {Soldevila}, \citenamefont {Tajahuerce}, \citenamefont {Gigan},
  \citenamefont {Rigneault},\ and\ \citenamefont {de~Aguiar}}]{Sturm2018}%
  \BibitemOpen
  \bibfield  {author} {\bibinfo {author} {\bibfnamefont {Benneth}\ \bibnamefont
  {Sturm}}, \bibinfo {author} {\bibfnamefont {Fernando}\ \bibnamefont
  {Soldevila}}, \bibinfo {author} {\bibfnamefont {Enrique}\ \bibnamefont
  {Tajahuerce}}, \bibinfo {author} {\bibfnamefont {Sylvain}\ \bibnamefont
  {Gigan}}, \bibinfo {author} {\bibfnamefont {Herve}\ \bibnamefont
  {Rigneault}}, \ and\ \bibinfo {author} {\bibfnamefont {Hilton~B.}\
  \bibnamefont {de~Aguiar}},\ }\bibfield  {title} {\enquote {\bibinfo {title}
  {High-sensitivity high-speed compressive spectrometer for raman imaging},}\
  }\href@noop {} {\bibfield  {journal} {\bibinfo  {journal} {arXiv:1811.06954}\
  } (\bibinfo {year} {2018})}\BibitemShut {NoStop}%
\bibitem [{\citenamefont {Xu}\ \emph {et~al.}(2012)\citenamefont {Xu},
  \citenamefont {Yin}, \citenamefont {Wen},\ and\ \citenamefont
  {Zhang}}]{Xu2012}%
  \BibitemOpen
  \bibfield  {author} {\bibinfo {author} {\bibfnamefont {Yangyang}\
  \bibnamefont {Xu}}, \bibinfo {author} {\bibfnamefont {Wotao}\ \bibnamefont
  {Yin}}, \bibinfo {author} {\bibfnamefont {Zaiwen}\ \bibnamefont {Wen}}, \
  and\ \bibinfo {author} {\bibfnamefont {Yin}\ \bibnamefont {Zhang}},\
  }\bibfield  {title} {\enquote {\bibinfo {title} {An alternating direction
  algorithm for matrix completion with nonnegative factors},}\ }\href@noop {}
  {\bibfield  {journal} {\bibinfo  {journal} {Frontiers of Mathematics in
  China}\ }\textbf {\bibinfo {volume} {7}},\ \bibinfo {pages} {365--384}
  (\bibinfo {year} {2012})}\BibitemShut {NoStop}%
\bibitem [{\citenamefont {Berto}\ \emph {et~al.}(2017)\citenamefont {Berto},
  \citenamefont {Scotté}, \citenamefont {Galland}, \citenamefont {Rigneault},\
  and\ \citenamefont {de~Aguiar}}]{Berto2017}%
  \BibitemOpen
  \bibfield  {author} {\bibinfo {author} {\bibfnamefont {P.}~\bibnamefont
  {Berto}}, \bibinfo {author} {\bibfnamefont {C.}~\bibnamefont {Scotté}},
  \bibinfo {author} {\bibfnamefont {F.}~\bibnamefont {Galland}}, \bibinfo
  {author} {\bibfnamefont {H.}~\bibnamefont {Rigneault}}, \ and\ \bibinfo
  {author} {\bibfnamefont {H.~B.}\ \bibnamefont {de~Aguiar}},\ }\bibfield
  {title} {\enquote {\bibinfo {title} {Programmable single-pixel-based
  broadband stimulated raman scattering},}\ }\href@noop {} {\bibfield
  {journal} {\bibinfo  {journal} {Opt. Lett.}\ }\textbf {\bibinfo {volume}
  {42}},\ \bibinfo {pages} {1696--1699} (\bibinfo {year} {2017})}\BibitemShut
  {NoStop}%
\bibitem [{\citenamefont {Krafft}\ \emph {et~al.}(2005)\citenamefont {Krafft},
  \citenamefont {Knetschke}, \citenamefont {Funk},\ and\ \citenamefont
  {Salzer}}]{Krafft2005}%
  \BibitemOpen
  \bibfield  {author} {\bibinfo {author} {\bibfnamefont {Christoph}\
  \bibnamefont {Krafft}}, \bibinfo {author} {\bibfnamefont {Thomas}\
  \bibnamefont {Knetschke}}, \bibinfo {author} {\bibfnamefont {Richard~HW}\
  \bibnamefont {Funk}}, \ and\ \bibinfo {author} {\bibfnamefont {Reiner}\
  \bibnamefont {Salzer}},\ }\bibfield  {title} {\enquote {\bibinfo {title}
  {Identification of organelles and vesicles in single cells by raman
  microspectroscopic mapping},}\ }\href@noop {} {\bibfield  {journal} {\bibinfo
   {journal} {Vibrational Spectroscopy}\ }\textbf {\bibinfo {volume} {38}},\
  \bibinfo {pages} {85--93} (\bibinfo {year} {2005})}\BibitemShut {NoStop}%
\bibitem [{\citenamefont {Donaldson~Jr}\ and\ \citenamefont
  {de~Aguiar}(2018)}]{Donaldson2018}%
  \BibitemOpen
  \bibfield  {author} {\bibinfo {author} {\bibfnamefont {Stephen~H}\
  \bibnamefont {Donaldson~Jr}}\ and\ \bibinfo {author} {\bibfnamefont
  {Hilton~B}\ \bibnamefont {de~Aguiar}},\ }\bibfield  {title} {\enquote
  {\bibinfo {title} {Molecular imaging of cholesterol and lipid distributions
  in model membranes},}\ }\href@noop {} {\bibfield  {journal} {\bibinfo
  {journal} {Journal of Physical Chemistry Letters}\ }\textbf {\bibinfo
  {volume} {9}},\ \bibinfo {pages} {1528--1533} (\bibinfo {year}
  {2018})}\BibitemShut {NoStop}%
\bibitem [{\citenamefont {Cai}\ \emph {et~al.}(2010)\citenamefont {Cai},
  \citenamefont {Cand{\`e}s},\ and\ \citenamefont {Shen}}]{Cai2010}%
  \BibitemOpen
  \bibfield  {author} {\bibinfo {author} {\bibfnamefont {Jian-Feng}\
  \bibnamefont {Cai}}, \bibinfo {author} {\bibfnamefont {Emmanuel~J}\
  \bibnamefont {Cand{\`e}s}}, \ and\ \bibinfo {author} {\bibfnamefont {Zuowei}\
  \bibnamefont {Shen}},\ }\bibfield  {title} {\enquote {\bibinfo {title} {A
  singular value thresholding algorithm for matrix completion},}\ }\href@noop
  {} {\bibfield  {journal} {\bibinfo  {journal} {SIAM Journal Optimization}\
  }\textbf {\bibinfo {volume} {20}},\ \bibinfo {pages} {1956--1982} (\bibinfo
  {year} {2010})}\BibitemShut {NoStop}%
\end{thebibliography}
%

\end{document}